\documentclass[12pt]{article}

\begin{document}
\rightline{RUB-TPII-16/98}
\begin{center}
{\Large
On the heavy quark mass expansion for the
operator $\bar Q\gamma_5Q$ and the charm content of $\eta, \eta'$ }\\
\vspace{0.4cm}
{\bf M. Franz$^a$, P.V. Pobylitsa$^{a,b}$, M.V. Polyakov$^{a,b}$,
K. Goeke$^{a}$}\\
\vspace{0.4cm}
{\it
$^a$Institut f\"ur Theoretische Physik II, Ruhr--Universit\"at Bochum,\\
 D--44780 Bochum, Germany\\
$^b$Petersburg Nuclear Physics Institute, 188350 Gatchina, Russia}
\end{center}
\vspace{0.1cm}

\begin{abstract}
 Recently in the context of studies
of the intrinsic charm content of the nucleon and of
the $\eta ^{\prime }$ meson two groups have arrived at different results for
the $1/m^3$ term of the heavy quark expansion for operator $\bar Q\gamma _5Q$
differing by the factor of six. We show that the form of both results
violates certain general conditions. Using the expression for the
axial anomaly with the finite Pauli-Villars regularization we obtain a new
expression for $1/m^3$ term of the heavy quark expansion for $\bar Q\gamma_5
Q$. With this new result we obtain an estimate for the constant
$f_{\eta'}^{(c)}\approx -\frac{1}{12 m_c^2}\langle0|
\frac{\alpha_s}{4\pi} G_{\mu\nu}^a \tilde
G^{\mu\nu,a}|\eta'\rangle \approx-2$~MeV
\end{abstract}
\vspace{0.3cm}

\noindent
{\bf 1.} The analysis of the charm content in noncharmed hadrons often 
appeals to the heavy quark expansion of operators like $\bar Q\gamma 
_5Q$ \cite{old,new,mog}.  One starts with the calculation of the 
expectation value of $ \langle \bar Q\gamma _5Q\rangle $ in the 
background gluon field $A$ with the subsequent heavy quark expansion in 
inverse powers of $m$ (we present the euclidean version)

\begin{equation}
\langle \bar Q\gamma _5Q(x)\rangle _A=\mbox{Sp}\langle x|\frac 1{\gamma _\mu
\nabla _\mu (A)+m}\gamma _5|x\rangle =\frac{c_1(x)}m+\frac{c_2(x)}{m^3}
+\ldots  \label{c-expansion}
\end{equation}
Recently two groups \cite{old} and \cite{new} came up with different results
for the coefficient $c_2$ differing by the factor of six. Specifically the
result of \cite{old} for the large $m$ expansion (\ref{c-expansion}) is
\begin{equation}
2m\mbox{Sp}\langle x|\frac 1{\gamma _\mu \nabla _\mu (A)+m}\gamma
_5|x\rangle =\frac 1{16\pi ^2}F_{\mu \nu }^a\tilde F_{\mu \nu }^a-\frac
1{16\pi ^2m^2}f_{abc}F_{\mu \nu }^a\tilde F_{\nu \alpha }^bF_{\alpha \mu
}^c+\ldots  \label{Toki}
\end{equation}
whereas the authors of \cite{new} obtain
\begin{equation}
2m\mbox{Sp}\langle x|\frac 1{\gamma _\mu \nabla _\mu (A)+m}\gamma
_5|x\rangle =\frac 1{16\pi ^2}F_{\mu \nu }^a\tilde F_{\mu \nu }^a-\frac
1{96\pi ^2m^2}f_{abc}F_{\mu \nu }^a\tilde F_{\nu \alpha }^bF_{\alpha \mu
}^c+\ldots  \label{Zhitnitsky}
\end{equation}

However, a simple argument shows that both results are not consistent with
the general properties of the axial anomaly. Indeed, the diagonal
matrix element of the operator in the lhs of eqs. (\ref{Toki}),
(\ref{Zhitnitsky}) is nothing else but the anomalous divergence of the
axial current $j_\mu ^5$ of a \emph{massless} fermion with the heavy
$m$ fermion considered as a Pauli--Villars \emph{regulator} (see e.g.
the review \cite{Morozov})

\begin{equation}
\left\langle \partial _\mu
j_\mu ^5(x)\right\rangle =2m\mbox{Sp}\langle x|\frac 1{\gamma _\mu
\nabla _\mu (A)+m}\gamma _5|x\rangle \; .
\label{anomaly}
\end{equation}

Eq. (\ref{anomaly}) is exact in the Pauli--Villars regularization and should
be valid in any order of the large $m$ expansion. On other hand, one can
write the local large $m$ expansion for the axial current in the background
gluon field
\begin{equation}
\left\langle j_\mu ^5(x)\right\rangle =\sum\limits_n\frac 1{m^n}X_n^\mu (x)
\; .
\end{equation}
This means that in the large $m$ expansion
\begin{equation}
\mbox{Sp}\langle x|\frac 1{\gamma ^\mu \nabla _\mu (A)+m}\gamma _5|x\rangle
=\sum\limits_n\frac 1{m^n}\partial _\mu X_n^\mu (x)\; .
\label{expansion-derivatives}
\end{equation}

However, the term $f_{abc}F_{\mu \nu }^a\tilde F_{\nu \alpha
}^bF_{\alpha \mu }^c$ in the rhs of the large $m$ expansions (\ref{Toki}),
(\ref{Zhitnitsky}) cannot be represented as a total derivative of a local
expression. Indeed, a straightforward calculation e.g. for the instanton
field shows that
\begin{equation}
\int d^4xf_{abc}F_{\mu \nu }^a\tilde F_{\nu \alpha }^bF_{\alpha \mu }^c\ne 0
\;.
\end{equation}
A simple dimensional
analysis shows that this nonvanishing contribution cannot come from the
surface terms if the instanton field is taken in the regular gauge.
Hence the integrand cannot be a total derivative.

We have recomputed the $1/m^3$ term of the expansion (\ref{c-expansion}) and
obtained the following result

\begin{equation}
2m\mbox{Sp}\langle x|\frac 1{\gamma ^\mu \nabla _\mu (A)+m}\gamma
_5|x\rangle =\frac 1{8\pi ^2}\mbox{Sp}_{{\rm color}}\left( F_{\lambda \nu
}\tilde F_{\lambda \nu }\right) +\frac 1{96\pi ^2m^2}\partial _\mu R_\mu
\; ,
\label{theres}
\end{equation}
\begin{equation}
R_\mu =\partial _\mu \mbox{Sp}_{{\rm color}}\left( F_{\lambda \nu }\tilde
F_{\lambda \nu }\right) -4\mbox{Sp}_{{\rm color}}\left( [\nabla _\alpha
,F_{\nu \alpha }]\tilde F_{\mu \nu }\right)
\; .  \label{res-2}
\end{equation}
Here the fundamental representation of the gauge group is implied
$F_{\lambda \nu }=F_{\lambda \nu }^a\frac{\lambda ^a}2$.
This expression differs from both earlier results (\ref{Toki}), (\ref
{Zhitnitsky}) and explicitly has the form of a total derivative in agreement
with the general form of the $1/m$ expansion (\ref{expansion-derivatives}).
In the case of the instanton field satisfying the self-duality
equation $F_{\lambda \nu }=\tilde F_{\lambda \nu } \; ,$
we find
\begin{equation}
\left. R_\mu \right| _{{\rm instanton}}=\partial _\mu
\mbox{Sp}_{{\rm color}}\left( F_{\lambda \nu }F_{\lambda \nu }\right)
=\partial _\mu \frac{96\rho
^4}{(x^2+\rho ^2)^4}\; ,
\end{equation}
where $\rho $ is the size of the instanton.

In principle the result (\ref{theres}) can be also extracted from the
calculation of the axial anomaly in the so-called
finite-mode regularization in \cite{Andrianov}.
Note that in that paper the parameter $m$ played the role
of the infrared regularization. Taking the ultraviolet cutoff of the
finite mode regularization to infinity one obtains from eq. (5.7) in
\cite{Andrianov} the following expression

\[
2m\mbox{Sp}\langle x|\frac 1{\gamma ^\mu \nabla _\mu (A)+m}\gamma
_5|x\rangle =\frac 1{16\pi ^2}\epsilon _{\mu \nu \lambda \rho }\mbox{Sp}_{
{\rm color}}(F_{\mu \nu }F_{\lambda \rho })+\frac 1{96\pi ^2m^2}\epsilon
_{\mu \nu \lambda \rho }
\]
\begin{equation}
\times \mbox{Sp}_{{\rm color}}\left\{ 2[\nabla _\alpha ,[\nabla _\alpha
,F_{\mu \nu }]]F_{\lambda \rho }+[\nabla _\alpha ,F_{\mu \nu }][\nabla
_\alpha ,F_{\lambda \rho }]-4iF_{\lambda \mu }F_{\nu \alpha }F_{\alpha \rho
}\right\}\;.  \label{res-1}
\end{equation}

After some simple algebraic transformations using the Bianchi identity,
expression (\ref{res-1}) can be brought to the form (\ref{theres}).
\vspace{0.2cm}

\noindent
{\bf 2.} Using the result (\ref{theres}) we can easily estimate the
matrix element:
\begin{equation}
\langle 0| \bar c \gamma_\mu \gamma_5 c|\eta'(q)\rangle=
i f_{\eta'}^{(c)} q_\mu\; .
\end{equation}
Combining the anomaly equation

\begin{equation}
\partial^\mu \bar c \gamma_\mu \gamma_5 c=
2 i m_c \bar c \gamma_5 c+
\frac{\alpha_s}{4\pi} G_{\mu\nu}^a \tilde G^{\mu\nu,a}\; ,
\label{anomalye}
\end{equation}
and our result (\ref{theres}) (rotated to the Minkowski space)
we obtain the folowing estimate of the constant $f_{\eta'}^{(c)}$
in the order $m_c^{-2}$:
\begin{equation}
f_{\eta'}^{(c)}=
-\frac{1}{12 m_c^2}
\langle0|
\frac{\alpha_s}{4\pi} G_{\mu\nu}^a \tilde G^{\mu\nu,a}|\eta'\rangle
\; .
\label{fc}
\end{equation}
In this equation we use the ``perturbative''
normalization for the gluon field strength
$ G_{\mu\nu}^a= F_{\mu\nu}^a/g$
and  the term proportional to
$\lbrack \nabla _\alpha ,G_{\nu \alpha }]$, which vanishes in pure
Yang-Mills theory, is neglected.
The omitted term with the help of QCD equation of motion
can be related to the following matrix element
\begin{equation}
\frac{\alpha_s}{4\pi}
\langle0|
g\, \sum_{f=u,d,s}
\bar \psi_f\gamma_\nu \tilde G_{\mu}^{\nu} \psi_f |\eta'\rangle \; .
\label{sfc}
\end{equation}
A method of computation of such matrix elements
in the framework
of the instanton model of the QCD vacuum was developed in ref.~\cite{DPW95}.
This method has already been applied to calculations of the
nucleon matrix elements of certain mixed quark-gluon operators
related to higher--twist corrections to deep-inelastic
scattering \cite{BPW97}. Rough order of maginitude estimate using
the results of ref.~\cite{BPW97} indicates that the omitted term
contributes (by absolute value)
at the level of $0.3$~MeV to the constant $f_{\eta'}^{(c)}$.
 Full quantitative estimates of the matrix element
(\ref{sfc}) will be given elsewhere.

Using the value
\begin{equation}
\langle0|
\frac{\alpha_s}{4\pi} G_{\mu\nu}^a \tilde G_{\mu\nu}^a|\eta'\rangle
= 0.056\; \mbox{GeV}^3\; ,
\end{equation}
obtained in \cite{DE}, we get from (\ref{fc}) the following estimate:
\begin{equation}
f_{\eta'}^{(c)}\approx - 2\; \mbox{MeV}
\; .
\end{equation}
Analogous estimate for the constant $f_{\eta}^{(c)}$ gives
$f_{\eta}^{(c)}\approx - 0.7\; \mbox{MeV}$,
if one uses the value
$\langle0|
\frac{\alpha_s}{4\pi} G_{\mu\nu}^a \tilde G_{\mu\nu}^a|\eta\rangle
= 0.020\; \mbox{GeV}^3$
obtained in ref.~\cite{DE}. Let us note
however that in the case of $\eta$ the contribution of the omitted
term can be of the same order, hence this estimate
should be considered as pure.
The full quantitative estimate of $f_{\eta'}^{(c)}$ and $f_{\eta}^{(c)}$
in the order $1/m_c^2$ of heavy quark mass expansion will be given
elsewhere.

To summarize, our result for the $m^{-3}$ term of the large $m$
expansion (\ref{c-expansion}) agrees with the general properties of the
axial anomaly in contrast to the expressions derived in \cite{old} and \cite
{new}. It should be noted that the expressions of both \cite{old} and
\cite{new}
were used by a number of authors for the analysis of the charm content of
noncharmed hadrons (see e.g. \cite{old,new,mog}).
The structure of our result (\ref{res-2}) essentially
differs from that of \cite{old} and \cite{new} which means that the results
of papers relying on \cite{old} and \cite{new} must be reconsidered.

\section*{Acknowledgments}
We are grateful to C. Weiss for careful reading the manuscript and
interesting remarks.
This work has been supported in part by a joint grant of the Russian
Foundation for Basic Research (RFBR) and the Deutsche
Forschungsgemeinschaft (DFG) 436 RUS 113/181/0 (R), by the BMBF grant
RUS-658-97, the COSY (J\"ulich) and the RFBR (Moscow). K.G. acknowledges
the hospitality of TRIUMF,Vancouver, and of the Special Research
Centre for the Subatomic Matter, Adelaide.  He thanks H.Fearing and
A.W.Thomas for useful discussions.

\end{document}